Exploring Elastic Net and Multivariate Regression

Matthias Raess

Ball State University

## Introduction

While cardiovascular disease and other non-communicable diseases, such as cancer, and diabetes, seem to run rampant in developed countries with the United States leading the way, followed by many European countries coming in at a close but almost negligible second, developing countries not only have to deal with communicable diseases, such as polio, tuberculosis, and malaria, but they are also often at a double disadvantage having to deal with communicable and non-communicable diseases (Boutayeb, 2006). Furthermore, it is a commonly known fact that the poor are usually hit hardest on a global scale, be it in developed or undeveloped countries (Gwatkin, Guillot, & Heuveline, 1999). Recent research has been focusing on how to prevent and counteract non-communicable diseases (Beaglehole et al., 2011; Kontis et al., 2014) while previous research focused on global and regional risk factors (Lopez, Mathers, Ezzati, Jamison, & Murray, 2006; Murray & Lopez, 1997), and still others have investigated the relationship between demographic factors and health and healthcare in general (Pol & Thomas, 2000).

It is clear from the amount of research that is being done in this field that we still do not know everything there is to know in terms of how demographic variables interact with health; thus the issue warrants more in-depth investigations on the subject. I will therefore devote this paper to the exploration of an unknown, extensive dataset in the attempt to shed light on some of the underlying and hidden relationships between several demographic variables and two of the health variables from the dataset.

Exploring Elastic Net and Multivariate Regression                                                3Hypothesis:

    $H_0$: Demographic information does not predict years of life lost due to communicable and non-communicable diseases.

    $H_A$: Demographic information predicts years of life lost due to communicable and non-communicable diseases.

## Methodology

**Dataset**

    The dataset originally comes from a blog (Canibais e Reis) which is now defunct. However, Valle-Jones (2010, March 9) used the dataset for a cluster analysis (PAM) of what the world eats; the R-code is available at the author's Github repository (Valle, 2010), and Huss and Westerberg (2010, March 15) describe the dataset and provide some caveats for using it. To get a better understanding of how nutrition, lifestyle, and health are related worldwide, the author combined information from the FAO Statistical Yearbook, the British Heart Foundation, the WHO Global Health Atlas, and the WHO Statistical Information System. These efforts resulted in a dataset with 101 variables on lifestyle, health, and nutrition from 86 countries (Huss & Westerberg, 2010, March 15). The most important caveat Huss and Westerberg (2010, March 15) make is to look out for confounding variables (they use PCA to work around that). Generally, the variables in the dataset fall into three broad categories: food variables containing information on dietary habits, demographic variables, and health variables.

**Pre-processing/regularization**

Initially, the dataset was split into three subsets (health, food, and demographic information), and, as a pre-processing step and to counteract confounding variables and spurious relationships, several variables were removed: summary variables such as meat in kcal/day (total) were included while specific meat types (e.g. bovine meat) were excluded for simplicity's sake, conceptual relationship, and to rule out collinearity. Further, I removed variables that did not have a clear frame of references as to what their original scales were. Variables with different years were also removed as well as variables that contain ambiguous information on gender and age. The remaining variables in the three subsets were then centered and standardized (z-scores). The problem inherent in the dataset at hand is that a relatively low number of cases (N=86) is met with over 100 variables, which need to be selected in such a way as to produce meaningful and significant results. In order to circumvent running a model with potentially highly correlated variables and to forgo a classical, and somewhat tedious, variable selection process (e.g. model comparison – AIC), I opt for the Elastic Net, a hybrid penalization method – first proposed by Zou and Hastie (2005) – which, as a penalization method, lies in between the LASSO and ridge regression. The Elastic Net (Zou & Hastie, 2005), was used to reduce the number of variables in the Elastic Net model; the reduced dataset is now ready for multivariate regression to find significant relationships between demographic information and health data. I opted against canonical correlation as an exploratory analysis tool because of the relatively small number of cases (countries, N=86), and, as Tabachnick and Fidell mention, one usually needs around ten cases/observations per variable in the model to get satisfying and statistically significant results (2013).

**Elastic net**

Proceeding with the pre-processed subsets, which now contain 21 variables (health), six variables (demographic information), and 32 variables (food) respectively, the Elastic Net was used as a penalization method; alpha was set to 0.5 halfway in between ridge regression, α = 0, and LASSO, α = 1 (Zou & Hastie, 2005). Alpha (α), according to Hastie and Qiang, is a higher level parameter, with which one needs to do some testing while lambda (λ) determines the overall strength of the penalty (Hastie & Qian, 2014). To address the hypothesis (relationship between demographic information and years of life lost due to communicable and non-communicable diseases), the first step is to build a model for the Elastic Net, which will then yield a reduced dataset for the following multivariate multiple regression with two correlated response variables.

Years of life lost to communicable diseases and years of life lost to non-communicable diseases serve as the two conceptually related outcome variables while the gross national income per capita, annual population growth rate, population in urban areas, total fertility rate per female, population with sustainable access to drinking water, and population with sustainable access to sanitation serve as the predictors. The analysis was run using the R *glmnet* package (Friedman, Hastie, & Tibshirani, 2010) with the *glmnet* function in RStudio (RStudio Team, 2016) and "mgaussian" as family for multivariate analysis.

## Results

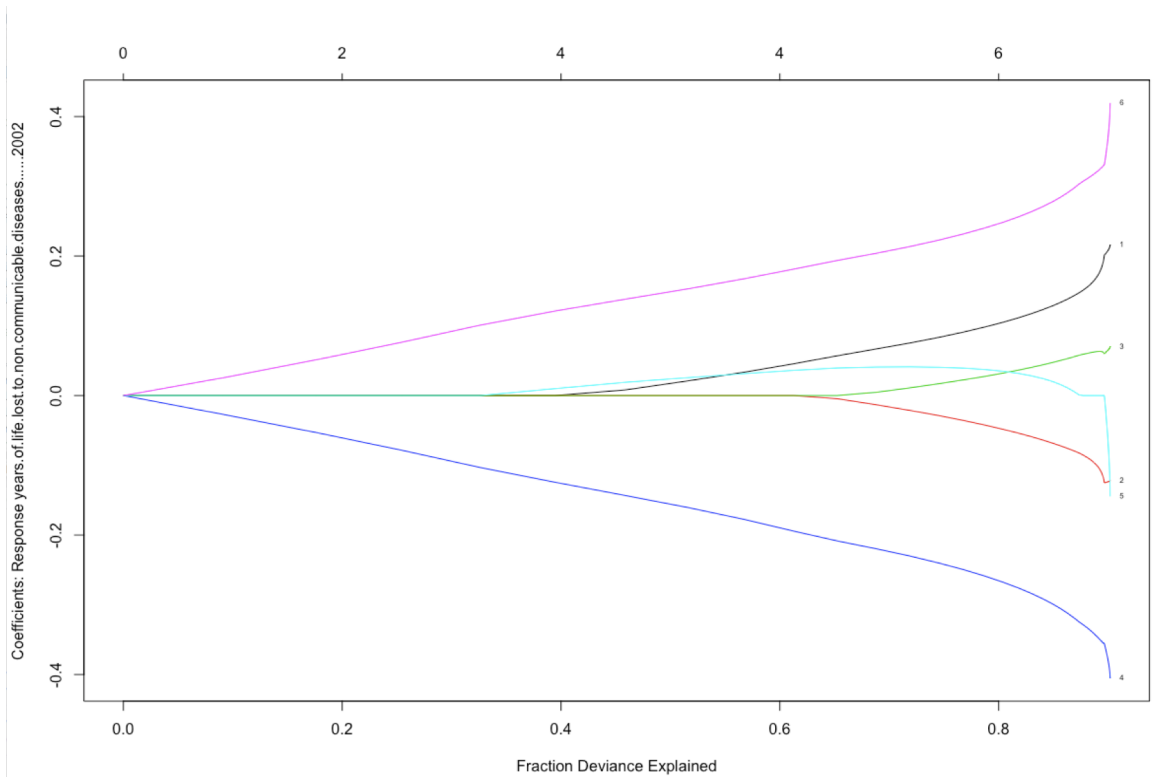

*Figure 1*. Fraction of deviance explained.

As can be inferred from Figure 1, the highest amount of deviance explained is in between four to six non-zero variables. In order to get the optimal lambda value, cross validation was conducted, which yielded *lambda.min* (0.033) as the optimal lambda value and the following coefficients for both response variables:

Table 1

*Elastic net coefficients*

|  | Years of life lost to communicable diseases | Years of life lost to non-communicable diseases |
|---|---|---|
| Fertility | 4.274144e-01 | -3.780597e-01 |
| Gross national income | -1.099295e-01 | 1.873516e-01 |
| Population annual growth | 3.833696e-02 | -9.896676e-02 |
| Population in urban areas | -9.147953e-02 | 5.414038e-02 |
| Population with access to improved drinking water | . removed | . removed |
| Population with access to sanitation | -3.807758e-01 | 3.377494e-01 |

The only variable that was dropped for both response variables is population with sustainable access to improved drinking water sources total. In order to double check for collinearity and to obtain the variance inflation factor (vif) for the predictor variables, two independent univariate multiple regressions were conducted. Significant regression equations were found in both cases ($F(5,80) = 156.8$, $p = < .001$; $R^2_{adj} = 0.90$, and $F(5,80) = 126$, $p < .001$; $R^2_{adj} = 0.88$). I chose not to report individual significance values for the independent variables here as the univariate regressions only serve to double-check collinearity. The only predictor to raise a concern in terms of the variance inflation factor was total fertility rate per female (vif = 4.53). However, it is generally assumed that a vif below 10 is not going to cause concern about collinearity between predictor variables – 4.53 is also still close to the more conservative cutoff, vif = 4, (Belsley, Kuh, & Welsch, 1980; O'Brien, 2007).

Table 2

*Multivariate multiple regression results*

|  | df | Approx F | num df | p |
|---|---|---|---|---|
| Fertility | 1 | 19.8432 | 2 | 1.041e-07 *** |
| Gross national income | 1 | 10.2582 |  | 0.0001095 *** |
| Population annual growth | 1 | 6.1968 | 2 | 0.0031630 ** |
| Population in urban areas | 1 | 4.4177 | 2 | 0.0151823 * |
| Population with access to sanitation | 1 | 19.4905 | 2 | 1.317e-07 *** |

Signif. codes: 0 '***' 0.001 '**' 0.01 '*'

The output of the multivariate multiple regression shows that all predictor variables have a significant relationship with year of life lost to communicable and non-communicable diseases. A look at the individual regression follow-ups (both significant, see also above; $F(5,80) = 156.8$, $p < .001$, and $F(5,80) = 126$, $p < .001$) reveals the following:

Table 3

*Univariate regression follow-ups*

|  | Years of life lost to communicable diseases | Years of life lost to non-communicable diseases |
|---|---|---|
| Fertility*** | .45*** | -.35*** |
| Gross national income*** | -.095$^{NS}$ | .22*** |
| Population annual growth** | .025$^{NS}$ | -.14* |
| Population in urban areas* | -.12* | .053$^{NS}$ |
| Population with access to sanitation*** | -.39*** | .33*** |

Signif. codes: 0 '***' 0.001 '**' 0.01 '*' not significant 'NS'

At this point, let us recall that all variables are centered and standardized (z-scores) and that years of life lost means the higher the number the more years of life lost. It is not surprising that the two outcome variables, which are conceptually related, do produce estimates with signs going in opposite directions (an initial Pearson's product moment correlation confirmed a strong inverse relationship, $r = -.98$, $p < 0.001$). While gross national income and annual population growth are not significantly related to years of life lost to communicable diseases, the other three variables have a significant relationship to the outcome variable: As the total fertility per female increases, so do the years of life lost to communicable diseases. As the population in urban areas increases, the years of life lost to communicable diseases goes down. We get a similar, even more significant, picture for population with access to sanitation (more civilization, more sanitation, fewer outbreaks of parasitic or bacterial diseases for instance). This is probably attributable to the improved overall infrastructure in more densely populated areas.

Ninety percent of the variance in years of life lost to communicable diseases is explained by the five predictor variables ($R^2_{adj} = 0.90$, $p < .001$). Looking at years of life lost to non-communicable diseases, we find an almost mirrored image. Higher fertility is negatively related to the outcome variable (years of life lost go down). The two non-significant variables, gross national income and population annual growth are significantly related to the outcome variable. Interestingly, as the gross national income goes up, so does the number of years of life lost due to non-communicable diseases. In an attempt to marry theory with practice, I conjecture that a more productive country yields higher gross income but also suffers more deaths due to heart attacks, cancer, and strokes. As the population grows, we find a negative relationship to the outcome variable. Again, the more access people have to sanitation (a sign for an industrialized economy), the years of life lost to non-communicable diseases increases - 88% of the variance in years of life lost to non-communicable diseases is explained by the five predictor variables ($R^2_{adj}$ = 0.88, $p$ < .001). In order to determine model fit, the most important assumptions are checked in the following:

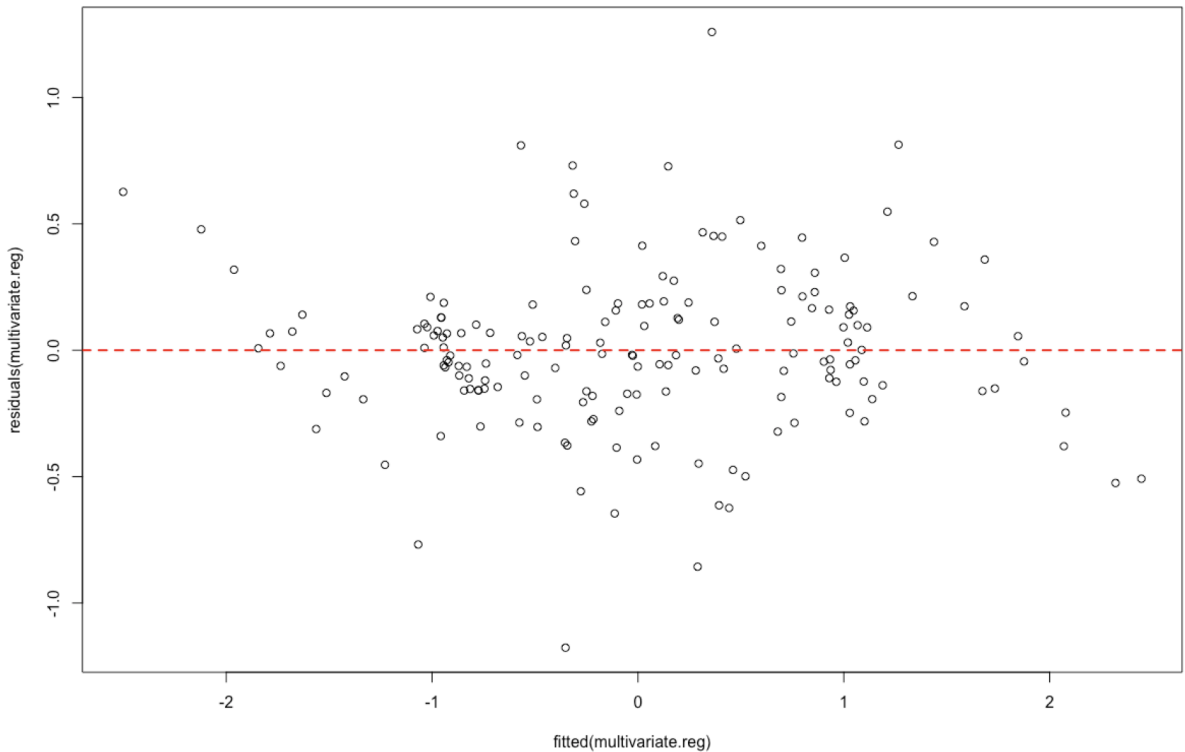

*Figure 2.* Residual plot for multivariate multiple regression

In terms of linearity, Figure 2 does not show a very distinctive pattern. However, the fitted values trail off on both ends of the spectrum, which could be an indicator of non-linearity and would ultimately need to be investigated further (a different model, such as a GAM, might have to be fitted in that case). Moving on to homoscedasticity, we can see that there is a blob-like structure in the middle of Figure 2 (which is what we want), again with both ends of the spectrum trailing off a bit – for the sake of argument, I am going to assume that both assumptions are met. The absence of collinearity has already been confirmed above (variance inflation factor), and since the assumption of normality of residuals is the least important, and linear models tend to be fairly robust against violations of normality, I am not going to go into detail about it here. Gellman and Hill (2007, p. 46) do not even recommend diagnostics on the normality assumption. While independence of observations, the assumption of assumptions, cannot really be checked, I assume that it is met because of the absence of multiple observations per country.

## Discussion

### Findings

The null hypothesis (demographic information does not predict years of life lost due to communicable and non-communicable diseases) is rejected at the alpha = .05 level for both univariate follow-ups to the multivariate regression, and the alternative hypothesis is adopted (demographic information predicts years of life lost due to communicable and non-communicable diseases). However, the interpretation of the results is not as straightforward as simply rejecting $H_0$.

I went into the analysis assuming that population growth would be significantly positively related to years of life lost to communicable diseases with a higher population density in mind, which plays a pronounced role in countries that lack good sanitation, such as India and Haiti – the opposite seems to be the case and population growth is not significantly related to years of life lost to communicable diseases. Furthermore, I did not expect fertility to be significantly related to both response variables – in Table 3 we can clearly see the mirror image: as fertility goes up so does the number of years of life lost to communicable diseases; conversely, it goes down for non-communicable diseases. I do not have a detailed, well investigated explanation for this relationship, but it is most likely attributable to latent factors not included in the model and the analysis. I did however expect gross national income to be significantly positively related to years of life lost to non-communicable diseases (e.g. higher standard of living, better healthcare system, and for reasons mentioned above). I also had a feeling that population in urban areas and population with continued access to sanitation would be significantly negatively related to years of life lost to communicable diseases – countries like India are probably an exception because the have some areas which are technologically developed but the overall situation of health, hygiene, and sanitation is sub-par (India has the highest amount of open defecation per square kilometer in the world (Berruti & Vyas, 2014)). Overall, these findings are interesting but largely expected and not really surprising in the grand scheme of things, but, nevertheless, confirm the hunch I initially had about some of the relationships. The size of the dataset and the large number of variables definitely calls for further investigation of different relationships, especially how health variables are related to food variables. As a matter of fact, there are maybe potentially more interesting things to be unearthed as for health and dietary information.

**Model, procedures, and dataset**

Both adjusted R-squared values are really high ($R^2_{adj}$= 0.90, $p < .001$, and $R^2_{adj}$= 0.88, $p < .001$ respectively), which either means the model(s) is/are really good or the increased number of predictors, or some other factor that I overlooked, contributed to overfitting the models. Possible solutions would include collecting more data, combining predictors, and shrinkage and penalization, which is what I have attempted (Babyak, 2004). Alternatively, it is possible that the skew in the multivariate distribution contributed to the inflation of R-squared and the model is not performing well. In addition, both residual standard errors are fairly low (0.31, and 0.34 respectively) and seem to indicate good model fit, which, however, could also be a sign of overfitting. Running univariate models twice seems redundant, but necessary if we want to get rid of collinearity especially in a scenario where no variable-reduction technique was used before running the multivariate regression. As mentioned above, a relatively new reduction technique like the Elastic Net is prone to be met with some issues: while trying to run a a hypothesis test to see which of the remaining variables were significantly correlated to the response variables, it turned out that the *selectiveInference* package in R (Tibshirani, Tibshirani, Taylor, Loftus, & Reid, 2016) does not yet accommodate multivariate testing, and so I moved on to run the multivariate regression with the reduced dataset. This however resulted in another error when I tried to use the R function, *vif*, which is part of the *car* package (Fox & Weisberg, 2011) to test for collinearity in the set of independent variables (to double check the elastic net result). I thus opted to run two separate univariate regression analyses to check collinearity and then do the multivariate regression.

These limitations notwithstanding, I attempted to show that a somewhat messy dataset (different scales, percentages, years, mg, etc.) that comprises over one hundred variables can be broken down into smaller conceptual subsets (food, health, demographic information), which then, can be further reduced by applying the Elastic Net. The result was a reduced dataset with 86 observations, two response variables, and five predictors. The R-code is available at the author's GitHub repository (Raess, 2016).